\documentclass[preprint,showpacs,preprintnumbers,amsmath,amssymb]{revtex4}

\usepackage{graphicx}
\usepackage{dcolumn}
\usepackage{bm}

\begin{document}

\preprint{APS/123-QED}

\title{The Role of Doppler Broadening in Electromagnetically Induced Transparency and Autler-Townes
Splitting in Open Molecular systems}

\author{
A.~Lazoudis,$^{\ntemple}$ E.~Ahmed,$^{\ntemple}$
L.~Li,$^{\nBeijing}$ T.~Kirova,$^{\ntemple}$ P.~Qi,$^{\ntemple}$
A.~Hansson,$^{\nStockholm}$ J.~Magnes,$^{\nWestPoint}$
F.C.~Spano,$^{\ntempleChem}$ and A.M.~Lyyra,$^{\ntemple}$ }
\affiliation{
\centerline{{$^{\ntemple}$Department of Physics, Temple
University, Philadelphia, Pennsylvania 19122}}
\centerline{{$^{\ntempleChem}$Department of Chemistry, Temple
University, Philadelphia, Pennsylvania 19122}}
\centerline{{$^{\nBeijing}$Department of Physics, Tsinghua
University, Beijing 100084, China}}
\centerline{{$^{\nStockholm}$Physics Department, Stockholm
University, Alba Nova, SE-10691 Stockholm, Sweden}}
\centerline{{$^{\nWestPoint}$Department of Physics, U.S Military
Academy, West Point, New York 10996-1790}}
}

\newcommand{\ntemple}{1}
\newcommand{\ntempleChem}{2}
\newcommand{\nBeijing}{3}
\newcommand{\nStockholm}{4}
\newcommand{\nWestPoint}{5}

\date{\today}

\begin{abstract}
We describe in this Letter how inhomogeneous line broadening
affects the Autler-Townes (AT) splitting in a three level open
molecular cascade system.  For moderate Rabi frequencies in the
range of 300 to 500~MHz the fluorescence line shape from the
uppermost level $|3\rangle$ in this system depends strongly on the
frequency ratio of the two laser fields. However, the fluorescence
spectrum of the intermediate level $|2\rangle$ appears as
expected.  We provide a description of the conditions for
optimally resolved AT splitting in terms of the probe
laser/coupling field frequency ratio and laser propagation
geometry based on our theoretical analysis of the Doppler
integral. This is important for applications such as molecular
angular momentum alignment as well as for the measurement of the
transition dipole moment matrix element.
\end{abstract}

\maketitle
Studies of coherence and quantum interference effects
\cite{1} such as Autler-Townes (AT) splitting and
electromagnetically induced transparency (EIT) in atomic and
molecular systems have attracted a great deal of attention,
because they yield a number of important potential applications
including production of slow light \cite{2}, lasing without
inversion \cite{3}, pulse matching effects \cite{4} and control of
the index of refraction \cite{5}. EIT was demonstrated for the
first time by Boller et al. \cite{6} in an atomic $\Lambda$-type
Strontium system. It renders an otherwise optically thick medium
transparent to a weak laser field. Most of the previous
theoretical and experimental work on EIT is based on various
three-level closed atomic systems with different energy level
configurations using both pulsed and continuous wave (cw) lasers
\cite{7,8,9,10,11}. To overcome inhomogeneous Doppler broadening
it is generally expected that high coupling field intensities are
required for observation of AT splitting and EIT. Nevertheless, it
has been shown that with proper selection of experimental
parameters and beam geometry it is possible to overcome this
difficulty and to demonstrate these effects in both atomic
\cite{8} and molecular systems \cite{12,13,Bergman} with coupling
field Rabi frequencies produced by commercially available cw
lasers.

In our previous papers \cite{12,13} we emphasized that the AT
effect can be used to directly measure the absolute value of the
molecular transition dipole moment as well as to facilitate
control of molecular angular momentum alignment. In this Letter we
investigate the role of the inhomogeneous line broadening and how
it affects the AT splitting in the fluorescence spectrum of levels
$|2\rangle$ and $|3\rangle$ in a 3-level open molecular cascade
system, when the coupling field is on resonance with the
$|2\rangle-|3\rangle$ rovibronic transition. We have observed that
the fluorescence line shape originating from the uppermost level
of the cascade excitation scheme critically depends on the probe
laser/coupling field wavenumber ratio, while at the same time the
fluorescence line shape observed from the intermediate level is
hardly affected at all. Based on our theoretical analysis to
explain these observations, we provide here a description of the
optimal conditions for AT splitting in terms of the coupling field
wavenumber compared to that of the probe laser as well as their
relative propagation direction.

\begin{figure}
\includegraphics{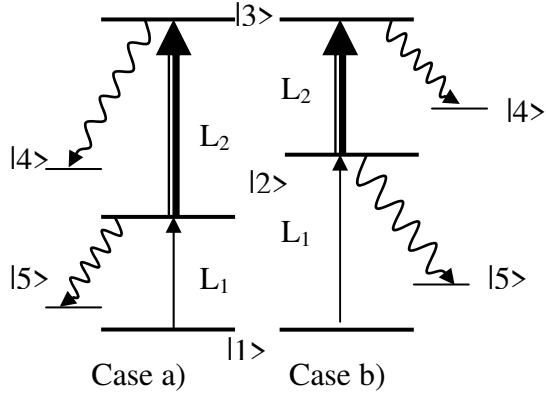}
\caption{\label{fig:fig1} The Na$_{2}$ cascade excitation scheme
for two different probe laser/coupling field wavenumber ratio $x =
k_{1}/k_{2}$, where k$_{1}$ is the weak probe laser (thin arrow)
and k$_{2}$ is the strong coupling field (thick arrow).  In case
(a) this ratio is smaller than 1 and in case (b) it is larger than
1. The energy levels chosen for case a) are: $|1\rangle$ =
$X^{1}\Sigma_{g}^{+}$(0,19), $|2\rangle$ =
$A^{1}\Sigma_{u}^{+}$(0,20), $|3\rangle$ = $2^{1}\Pi_{g}$(0,19),
$|4\rangle$ = $A^{1}\Sigma_{u}^{+}$(0,18) or
$A^{1}\Sigma_{u}^{+}$(0,20), and $|5\rangle$ =
$X^{1}\Sigma_{g}^{+}$(3,21). Those for case b) are: $|1\rangle$ =
$X^{1}\Sigma_{g}^{+}$(1,19), $|2\rangle$ =
$A^{1}\Sigma_{u}^{+}$(3,18), $|3\rangle$ =
4$^{1}\Sigma_{g}^{+}$(0,17),
$|4\rangle$=$A^{1}\Sigma_{u}^{+}$(1,16) and $|5\rangle$ =
$X^{1}\Sigma_{g}^{+}$(2,19). The laser frequencies for transitions
$|1\rangle$ to $|2\rangle$ and $|2\rangle$ to $|3\rangle$ are
L$_{1}$ (14647.547 cm$^{-1}$) and L$_{2}$ (15888.065 cm$^{-1}$)
for case a) and L$_{1}$ (14828.639 cm$^{-1}$) and L$_{2}$
(13284.554 cm$^{-1}$) for case b). Single channel side
fluorescence wavenumbers were 15771.91$\pm$0.3 cm$^{-1}$ and
14167.36$\pm$0.3 cm$^{-1}$ for case a) transitions $|3\rangle$ to
$|4\rangle$ and $|2\rangle$ to $|5\rangle$, respectively. These
values were 13522.93$\pm$0.3cm$^{-1}$ and
14672.59$\pm$0.3cm$^{-1}$ for case b). The lifetimes are 12.2ns
for level $|2\rangle$ in cases a) and b) \cite{14}, and 21.0ns and
12.7ns for level $|3\rangle$ in cases a) and b)
\cite{15,16,18,19}. }
\end{figure}

\begin{figure}
\includegraphics{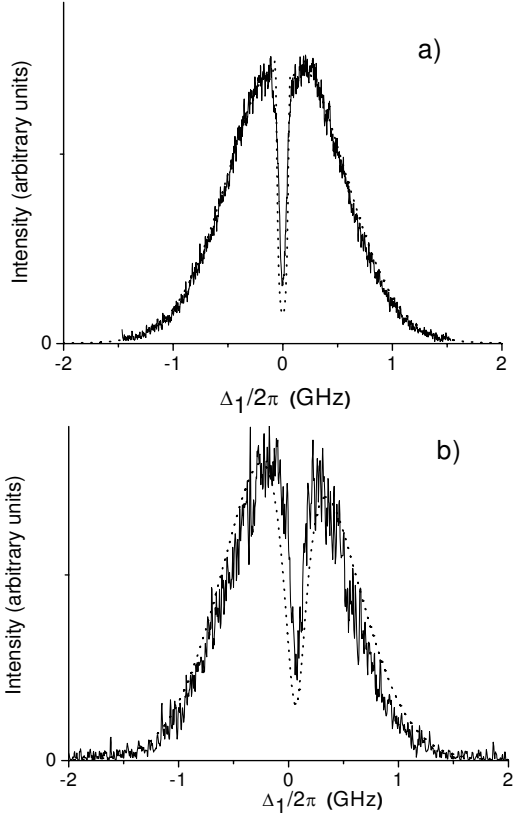}
\caption{\label{fig:fig2} Observation of EIT by fluorescence
detection from the intermediate level $|2\rangle$ as a function of
the probe laser detuning for counter-propagating beams with a)
$|k_{1}/k_{2}|$ = 0.922 and $\Delta_{2} = 0$ MHz and b)
$|k_{1}/k_{2}|$ = 1.116 and $\Delta_{2} = 60$ MHz.  The Doppler
width is 1.1GHz corresponding to 625K sample temperature. The
dotted line represents the simulations. }
\end{figure}

For the experimental set up adopted in our study we refer the
reader to \cite{12} and \cite{13}. Schematic illustrations of the
Na$_{2}$ energy levels for the cascade configuration discussed
here are displayed in the two panels of Fig.~\ref{fig:fig1} Case
a) illustrates a situation in which the absolute value of the
ratio $k_{1}/k_{2}$ of the probe laser wave number to that of the
coupling field is smaller than unity. We define this ratio as
positive for co-propagating beams and negative for the counter
propagating configuration. In case b) the absolute value of the
$k_{1}/k_{2}$ ratio is greater than one instead.
In addition to obeying this constraint on the transition
frequencies, the selected energy levels were chosen to allow the
generation of the largest Rabi frequency of the coupling field,
given the available laser sources [narrowband frequency stabilized
dye and Titanium Sapphire lasers (Coherent Autoscan 699-29 and
899-29)]. Intermediate level $|2\rangle$ belongs to the Na$_{2}$
$A^{1}\Sigma_{u}^{+}$ state and the uppermost level belongs to the
2$^{1}\Pi_{g}$ state in case a) and to the 4$^{1}\Sigma_{g}^{+}
$state in case b), respectively, as indicated in the caption of
Fig.~\ref{fig:fig1}, where, in addition, we provide the
fluorescence detection wavelengths in wavenumber units. With the
coupling laser tuned to resonance, we scanned the probe laser and
detected fluorescence from level $|2\rangle$ and from level
$|3\rangle$ by isolating the desired fluorescence from the
background radiation with a Spex 1404 double monochromator used as
a narrow band filter. In both cases a) and b) the fluorescence
from the intermediate level $|2\rangle$ displayed the EIT pattern
expected in an open molecular system \cite{12} as illustrated in
Figs.~\ref{fig:fig2} a) and b).
However, the AT splitting, which is generally present in the
fluorescence spectrum from level $|3\rangle$ for a sufficiently
large coupling field Rabi frequency, is quite sensitive to the
wavenumber ratio $x = k_{1}/k_{2}$. In case a) when the absolute
value of this ratio is smaller than unity, we observed a clear AT
splitting, as shown in Fig. 3a, whereas in case b) with
$|k_{1}/k_{2}| >1$, we did not observe evidence of this effect at
all (see Fig. 3b).The Rabi frequencies of the coupling field were
400MHz and 530MHz in case a) and b),respectively. Thus case b)
should be more favorable for the observation of AT splitting with
regard to this parameter.

\begin{figure}
\includegraphics{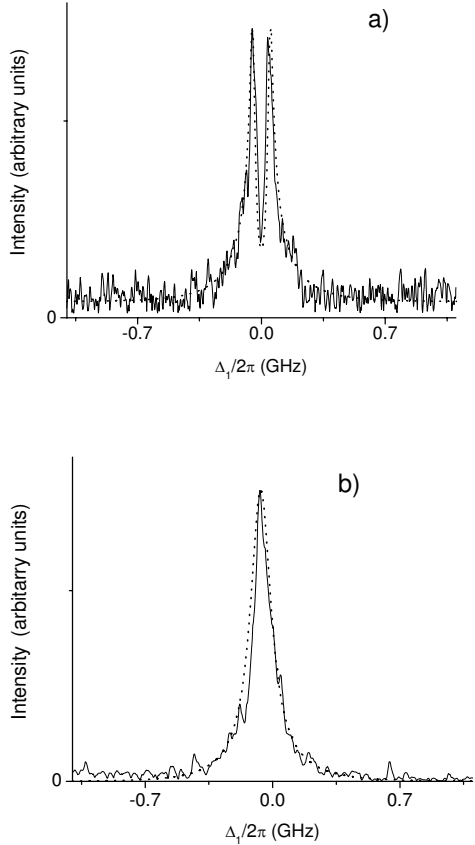}
\caption{\label{fig:fig3} Level $|3\rangle$ fluorescence under the
presence of a strong  coupling field for: a) $k_{1}/k_{2}$ = -
0.922 and b) $k_{1}/k_{2}$ = - 1.116. The dotted line represents
the simulations.}
\end{figure}

We simulated the experimental data by using stationary solutions
of the density matrix equations of motion as shown in
Figs.~\ref{fig:fig2} and~\ref{fig:fig3}.  Our Rabi frequencies
were 6 MHz and 400 MHz for the probe and coupling fields,
respectively, for case a) and 36 MHz and 530 MHz for case b).
These values were determined from the calculated transition dipole
moment matrix elements \cite{16}, Franck Condon factors for known
molecular potentials \cite{15,18,19} and the measured E field
amplitude in the interaction region \cite{knife}. Our results show
that the presence (or absence) of a splitting in the fluorescence
spectrum of the upper level $|3\rangle$ depends in a sensitive way
on the wavenumber ratio of the probe and coupling field lasers. As
we explain below, the origin of this result is the Doppler
broadening.

In order to gain physical insight into the behavior of the system
we used the perturbative model discussed by Qi et al. in Ref.
\cite{12} for an open molecular cascade system. We have
investigated the conditions under which the fluorescence spectrum
from level $|3\rangle$ displays the experimentally observed shown
behavior in Figs.~\ref{fig:fig3}a. and~\ref{fig:fig3}b. The
calculated Doppler averaged fluorescence spectrum from the upper
level $|3\rangle$ (summed over the magnetic sublevels of the
$|2\rangle \rightarrow|3\rangle$ transition), as a function of the
detunings $\Delta_{1} = \omega_{21} - \omega_{1} +
k_{1}\emph{v}_{z}$ and $\Delta_{2} = \omega_{32} - \omega_{2} +
k_{2}\emph{v}_{z}$ of the probe and coupling fields, respectively,
is given by
\begin{equation}
\emph{I}_{3}(\Delta_{1},\Delta_{2}) = \sum_{\text{M}}
\int_{-\infty}^{+\infty}\rho_{33}^{M}(\emph{v}_{z})\emph{N}(\emph{v}_{z})\emph{d}v_{z}
\;,
\label{eq:one}
\end{equation}
where z is the axis of propagation of the laser beams,
\emph{v}$_{z}$ is the z-component of the thermal velocity of the
molecules, \emph{N}(\emph{v}$_{z}$) is the Gaussian velocity
distribution, and $\rho_{33}$ is the analytical expression for the
population of level $|3\rangle$, given in Ref. 12.
We were able to carry out the integration of the Doppler integral
analytically. However, the details of this calculation with the
full result will be reported elsewhere \cite{20}, because of the
complexity of the mathematical expressions and the limited space
of the current manuscript. Based on this calculation the intensity
of the fluorescence spectrum $I_{3}(\Delta_{1},\Delta_{2})$ from
level $|3\rangle$ is proportional to the following factors:
\begin{equation}
\emph{I}_{3}(\Delta_{1},\Delta_{2})
\propto\frac{\emph{w}(\emph{z}_{i})}{\emph{z}_{1}-\emph{z}_{2}}
\;,
\label{eq:2}
\end{equation}
where $w(z_{i})$ is the Faddeeva function \cite{Fadd} and z$_{1}$
and z$_{2}$ are the roots of the denominator of
$\rho_{33}(\Delta_{1},\Delta_{2})$ from Ref. \cite{12}. Our
analysis indicates that the behavior of $I_{3}$ stems from the two
factors, $w(z_{i})$ and $(z_{1}-z_{2})^{-1}$.

\begin{figure}
\includegraphics{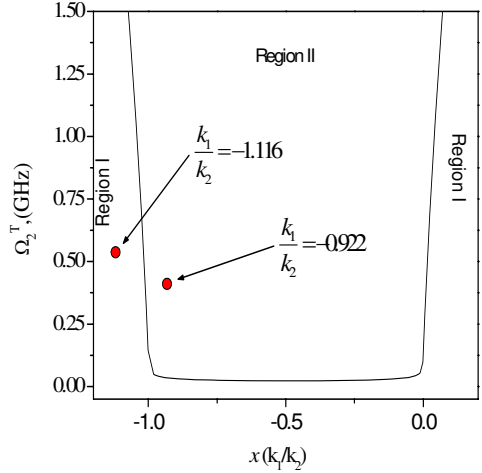}
\caption{\label{fig:fig4} Threshold Rabi frequency
$\Omega_{2}^{T}$ of the coupling field, as a function of the ratio
$k_{1}/k_{2}$. AT splitting can be observed only in region II. The
$\Omega_{2}^{T}$ value is much smaller for $-1<k_{1}/k_{2}<0$ than
for $-1>k_{1}/k_{2}$ or $k_{1}/k_{2}>0$ (co-propagating geometry).
The two dots indicate the $k_{1}/k_{2}$ ratio of our experiments.}
\end{figure}

\begin{figure}
\includegraphics{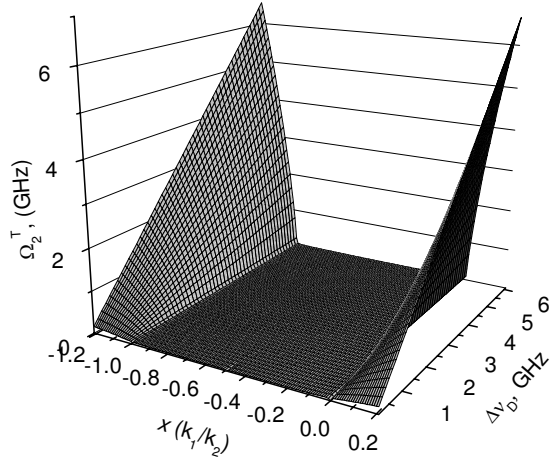}
\caption{\label{fig:fig5} Calculated threshold Rabi frequency
$\Omega_{2}^{T}$ as a function of the wavenumber ratio
$x=k_{1}/k_{2}$ and the Doppler linewidth $\Delta\nu_{D}$.
$\Omega_{2}^{T}$ does not depend on $\Delta\nu_{D}$ in the
interval $-1<k_{1}/k_{2}<0$, in contrast there is very strong
dependence in $-1>k_{1}/k_{2}$ or $k_{1}/k_{2}>0$. }
\end{figure}

Based on the analytically calculated
$I_{3}(\Delta_{1},\Delta_{2})$ we have derived expressions for the
threshold Rabi frequency $\Omega_{2}^{T}$ as a function of the
ratio $k_{1}/k_{2}$ above which AT splitting can be experimentally
observed.  For this we have used the condition for change in the
curvature of the $I_{3}(\Delta_{1},\Delta_{2})$ at zero detuning
of the probe laser:
\begin{equation}
\left\{\frac{\partial^{2}(\Delta_{1},..,\Omega_{2}^{M-1},\Omega_{2}^{M},\Omega_{2}^{M+1},..,x)}{\partial\Delta_{1}^{2}}\right\}_{\Delta_{1}=0}=0
\label{eq:3}
\end{equation}
The results are summarized in Fig.~\ref{fig:fig4}, which shows
that region II is most favorable for the observation of AT
splitting with the smallest required  coupling field strength.
This corresponds to a configuration in which the probe and
coupling laser beams counter-propagate and $|k_{1}/k_{2}|$ is
smaller than unity. The dots in the figure identify the values of
the $k_{1}/k_{2}$  ratio that were used in our experiments.
We have also investigated the dependence of this threshold
$\Omega_{2}^{T}$ on the inhomogeneous Doppler linewidth
$\Delta\nu_{D}$. The 3 dimensional plot of Fig.~\ref{fig:fig5}
shows a simulation with the same parameters as Fig.~\ref{fig:fig4}
where the 3$^{rd}$ dimension is represented by $\Delta\nu_{D}$.
The threshold Rabi frequency  $\Omega_{2}^{T}$ grows very rapidly
with increasing Doppler linewidth in the region $k_{1}/k_{2}<-1$
and $k_{1}/k_{2}>0$ and has no $\Delta\nu_{D}$ dependence at all
in region $-1<k_{1}/k_{2}<0$.
This result is unexpected, and quite interesting. It shows that
only when the two beams counter propagate, and the coupling laser
has a larger wavenumber $k_{2}$ than the probe laser $k_{1}$, AT
splitting of the same magnitude can be observed independently of
the molecular/atomic weight or the temperature of the system. In
this region only a moderate coupling field Rabi frequency is
required to split the experimental Optical Optical Double
Resonance (OODR) spectrum and to observe the AT effect independent
of the Doppler linewidth $\Delta\nu_{D}$. In order to explain this
behavior we have to consider the expression for
$I_{3}(\Delta_{1},\Delta_{2})$ in Eq.2. First for $-1<x<0$ our
analysis \cite{20} shows that the splitting is mainly determined
by:
\begin{equation}
\frac{1}{\emph{z}_{1}-\emph{z}_{2}}=\frac{2\emph{x}(1+\emph{x})[(\Delta_{1}-\emph{i}\Gamma)^{2}+\emph{x}(1+\emph{x})\Omega_{2}^{2}]^{\frac{1}{2}}}
{[(\Delta_{1}^{2}-\Gamma^{2}+\emph{x}(1+\emph{x})\Omega_{2}^{2})^{2}+4\Gamma^{2}\Delta_{1}^{2}]^{\frac{1}{2}}}
\label{eq:4}
\end{equation}
where  $\Gamma$ is defined as $\Gamma = \gamma_{12}(1 + \emph{x})
- \gamma_{13}\emph{x}$.\\
This expression is independent of $\Delta\nu_{D}$. Therefore the
threshold $\Omega_{2}^{T}$ becomes independent of the Doppler
linewidth. This has an important physical implication in that the
AT splitting will be equally well resolved independent of the
magnitude of the Doppler broadening $\Delta\nu_{D}$. On the
contrary, in the region where the ratio is $x<-1$ or $x>0$ the AT
splitting and the threshold value of $\Omega_{2}^{T}$ are
determined by
\begin{equation}
\emph{w}(\emph{z}_{i})\propto\emph{e}^{-\emph{z}_{\emph{i}}^{2}}\propto\emph{e}^{-\frac{f(\Omega_{2},x,\gamma_{ij})}{(\Delta\nu_{D})^{2}}}
 \label{eq:5}
\end{equation}
for sufficiently large $|z_{\emph{i}}|$, indicating a strong
dependence on $\Delta\nu_{D}$. Thus the  increase in the value of
$\Delta\nu_{D}$ leads to a rapid increase in the threshold Rabi
frequency $\Omega_{2}^{T}$. This explains why the experimental
spectrum in Fig.~\ref{fig:fig3} does not show AT splitting, since
our available coupling Rabi frequency was 300$\sim$350MHz, while
the required value is of the order of a GHz, for this particular
experiment.

In conclusion, our experimental results show profoundly different
behavior of the AT splitting in the spectra from the upper level
in a Doppler broadened open molecular cascade system, while the
fluorescence spectrum emanating from the intermediate level is
almost unaffected. We have shown that this unexpected behavior is
based on the presence of Doppler broadening and that it is
strongly dependent on the ratio of the probe and coupling laser
wavenumbers. By analyzing the dependence of the threshold Rabi
frequency $\Omega_{2}^{T}$ on the probe laser-coupling field
wavenumber ratio and the Doppler linewidth $\Delta\nu_{D}$ we have
provided an overview of the optimal conditions for observing AT
splitting in this excitation scheme.

\begin{acknowledgments}
This work was supported by National Science Foundation Awards PHY
0245311 and PHY 0216187. We gratefully acknowledge valuable
discussions with Prof. Lorenzo M. Narducci of Drexel University.
\end{acknowledgments}

\end{document}